\documentclass[openacc]{rsproca_new} 
\usepackage{bm}
\usepackage{siunitx}
\usepackage{tikz}
\usepackage{tabularx}

\usepackage{xcolor}

\titlehead{Research}

\begin{document}

\title{Guided elastic waves informed material modelling of soft incompressible media}

\author{
Pierre Chantelot$^{1}$, Samuel Croquette$^{1}$ and Fabrice Lemoult$^{1}$}

\address{$^{1}$Institut Langevin, ESPCI Paris, Université PSL, CNRS, 75005 Paris}

\subject{xxxxx, xxxxx, xxxx}

\keywords{Soft matter, Acoustoelasticity, Viscoelasticity, Guided elastic waves}

\corres{Pierre Chantelot\\
\email{p.chantelot@gmail.com}}

\begin{abstract}
Identifying a universal material constitutive law, that describes the mechanical response of rubber-like solids for all deformation fields and achievable extensions, is an outstanding challenge.
Here, we propose to exploit the propagation of elastic waves and demonstrate that monitoring incremental guided wave propagation in an elastomer plate undergoing uniaxial extension reveals model sensitivities that are inaccessible in the corresponding static test.
We measure the dispersion relations of the three zero-order guided modes, propagating parallel and perpendicular to the direction of imposed elongation.
We compare them with predictions from the acoustoelastic theory, that also take into account material rheology, using parameters extracted from fitting the uniaxial stress-strain curve across three successive elongation regimes, following the methodical procedure of Destrade \emph{et al.}~\cite{destrade2017}.
We evidence that our approach lifts the degeneracy between hyperelastic models with different functional forms of the so-called $C_2$ term, which remain undistinguishable from static uniaxial tension stress–strain measurements alone.
However, like their static counterpart, our dynamics measurements cannot distinguish between different generalized neo-Hookean models.
\end{abstract}





\maketitle

\section{Introduction}
Natural rubber, synthetic elastomers and biological tissues possess the remarkable ability to sustain large deformation, and typically exhibit a complex, non-linear, mechanical response~\cite{treloar2005}.
Modelling and predicting the behaviour of such rubber-like solids is relevant for a wide range of applications such as biomechanics~\cite{goriely2005,holzapfel2010}, soft robotics~\cite{laschi2016,ng2021}, and the design of bio-inspired materials~\cite{mondal2020,li2022_2}.
Yet, finding a universal material constitutive law that can accurately describe the mechanical behaviour of a given elastomer for all possible deformation fields and for the whole range of achievable extensions remains an outstanding challenge~\cite{destrade2017}.

Recalling the origins of rubber elasticity helps in navigating the large number of constitutive models proposed in the literature.
Rubber-like solids are composed of long-chain molecules that interact through weak forces, and that are linked together to form a sparse network whose mechanical properties are of entropic nature~\cite{treloar2005}.
Using idealized chain models, it is possible to obtain the chain's spatial conformation and to determine the macroscopic network properties through a statistical averaging procedure.
Assuming the chains never approach their extended length, \emph{i.e.} using the Gaussian chain approximation, this micro-mechanical approach yields the neo-Hookean strain-energy density~\cite{treloar2005}.

Further developments addressed the limitations of the neo-Hookean theory evidenced by static experiments across multiple deformation classes~\cite{treloar1944}.
They followed two distinct paths.
(i) The statistical theory was refined by relaxing the assumptions of the Gaussian chain model, enabling the capture of the strain-stiffening effect at large elongation associated with finite chain extensibility~\cite{wang1952,arruda1993}.
(ii) Phenomenological continuum mechanics models were introduced to account for the experimental deviations from neo-Hookean predictions.
Most notably, Mooney and Rivlin proposed to augment the neo-Hookean constitutive law with an additional term, the so-called $C_2$ term~\cite{mooney1940,rivlin1948}, and Ogden suggested an empirical form of the strain-energy that allowed, for the first time, to fit a single model across a variety of deformation classes and a large range of strain~\cite{ogden1972}.
Rather than being mutually exclusive, these two approaches have proven to be complementary.
Non-Gaussian models have been connected to generalized neo-Hookean continuum models that capture strain-hardening~\cite{puglisi2016}.
For example, Horgan \& Saccomandi~\cite{horgan2002} evidenced the link between the Arruda-Boyce model~\cite{arruda1993} and the phenomenological Gent model~\cite{gent1996}.
And the additional term proposed by Mooney and Rivlin can also be linked to micro-mechanical considerations by modelling the constraints from neighbouring chains using a tube model~\cite{kroon2011,puglisi2016}, yielding a $C_2$ contribution of the form proposed by Gent \& Thomas~\cite{gent1958}.

Assessing the performance of the constitutive laws introduced to overcome the limitations of the neo-Hookean strain-energy requires determining their ability to fit experimental data for all deformation classes and all extensions.
However, Ogden \emph{et al.}~\cite{ogden2004} identified that the fitting procedure faces a fundamental challenge: multiple sets of parameters can give an equally good fit of uniaxial extension data, but markedly different predictions for other deformations or for simple boundary value problems.
Destrade \emph{et al.}~\cite{destrade2017} took a key step towards the resolution of this issue by delineating a methodical procedure to find a unique optimal parameter set when fitting simple extension data, the most commonly performed test to characterize rubber-like solids.
They evidenced that any standard linear combination of a generalized neo-Hookean model and a $C_2$ term captures the mechanical response in simple tension, highlighting the need to perform experiments across multiple deformation classes to discriminate constitutive models. 

Our aim here is to show that dynamic measurements provide an alternative approach.
We demonstrate how guided waves propagating in an elastomer plate undergoing uniaxial extension reveal model sensitivities that remain hidden in the corresponding static test.
We first describe the experiments from which we obtain the dispersion relations of the two lowest order in-plane modes and the lowest order out-of-plane mode propagating parallel and perpendicular to the direction of elongation (section \ref{sec:2}).
In section \ref{sec:3}, we recall the modelling of the mechanical response of nearly incompressible hyperelastic materials within the non-linear elasticity framework.
We then detail how we compare the dispersion relations of guided waves with predictions from the acoustoelastic theory, that depend on the parameters extracted from fitting the uniaxial stress-strain curve using the procedure of Destrade \emph{et al.}~\cite{destrade2017}, and highlight how guided modes carry additional insight on the constitutive law.
In section \ref{sec:4}, we present the results of this comparison distinguishing the three regimes introduced in~\cite{destrade2017}: the small to moderate elongation regime, the strain-hardening regime, and the large deformation regime.
We show that while multiple constitutive models yield indistinguishable fits of the uniaxial stress-strain curve, their predictions for wave dispersion diverge significantly.
We evidence that the dynamic measurements are sensitive to the form of the $C_2$ energy term, unlike the static response, but that any standard generalized neo-Hookean model captures the guided waves behaviour in the limiting chain regime.
Finally, we conclude by contrasting our observations with that gained by considering other deformations classes, and discuss the perspectives of this work in section \ref{sec:6}.
\section{Measuring guided elastic waves in a plate}
\label{sec:2}
We characterize wave propagation in a plate subjected to a large, static, uniaxial deformation where we impose the principal stretch $\lambda$ in the direction $\boldsymbol{e}_1$, as sketched in figure~\ref{fig:1}(a).
We focus on the three zero-order guided modes propagating in the plate: (i) the non-dispersive shear horizontal mode $S\!H_0$ which corresponds to shear waves polarized in the $(\boldsymbol{e}_1, \boldsymbol{e}_3)$ plane, and the two Lamb modes: (ii) the  $S_0$ mode which resembles a longitudinally polarized mode in the plane $(\boldsymbol{e}_1, \boldsymbol{e}_3)$ and (iii) the flexural $A_0$ mode which mostly exhibits out-of-plane displacement in the direction $\boldsymbol{e}_2$~\cite{royer1999}.  
We extend the experiments of Delory \emph{et al.}~\cite{delory2023} that obtained the dispersion relations of the in-plane $S\!H_0$ and $S_0$ modes propagating both parallel and perpendicular to the elongation direction for varying values of $\lambda$, by additionally measuring the dispersion curves of the out-of-plane $A_0$ mode.

\subsection{In-plane modes}
\begin{figure}
    \centering
    \includegraphics[width = \textwidth]{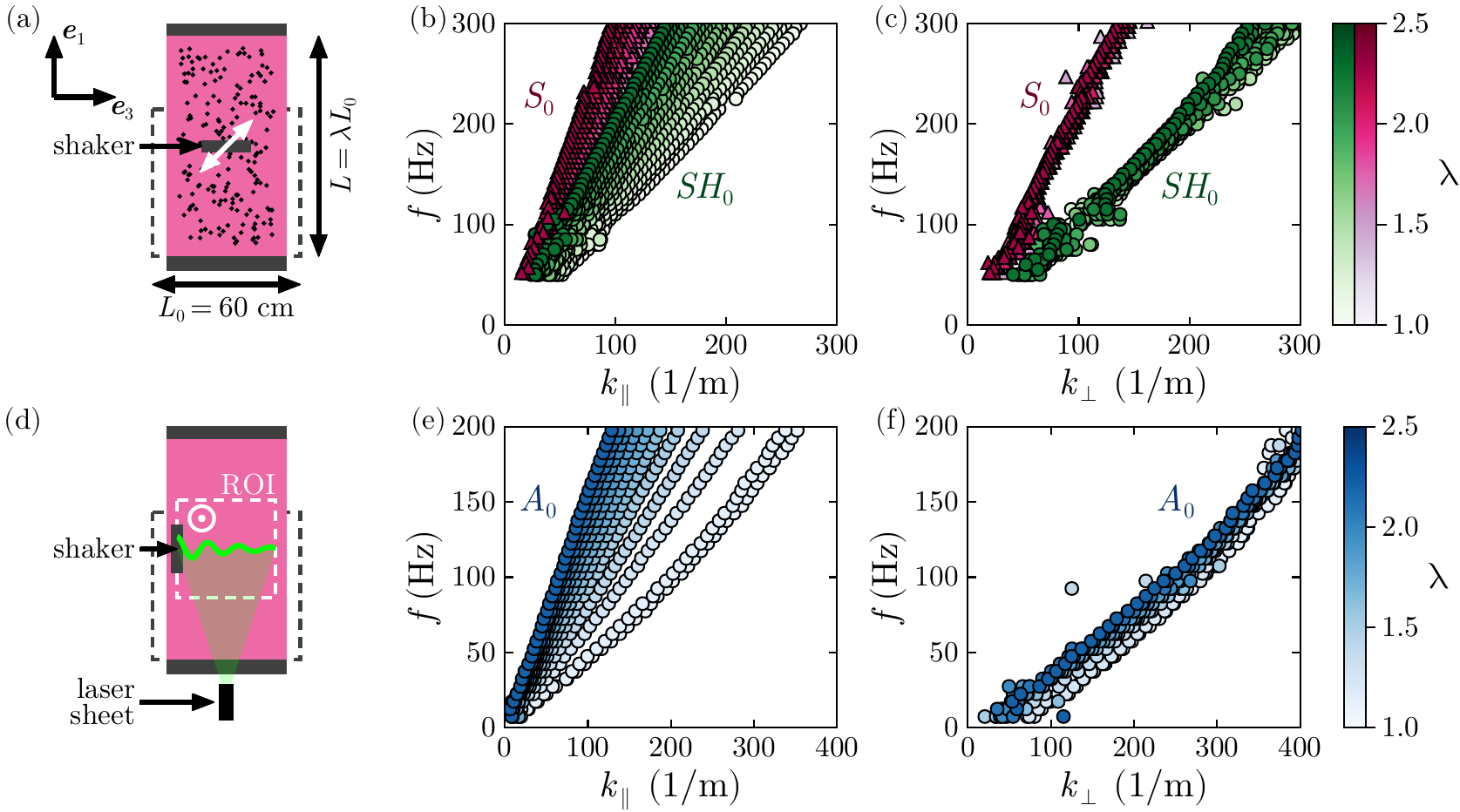}
    \caption{\label{fig:1} (a) Sketch of the experimental setup used to measure the dispersion curves of the in-plane modes in a soft elastomer plate.
    (b-c) Dispersion relations of the in-plane $S\!H_0$ (green circles) and $S_0$ (pink triangles) modes propagating parallel and perpendicular to the direction of elongation for $\lambda \in [1.03,2.27]$.
    (d) Schematic of experimental setup used to obtain the dispersion relation of the out-of-plane $A_0$ mode.
    (e-f) Dispersion curves of the $A_0$ mode propagating parallel and perpendicular to the imposed stretch for $\lambda \in [1.11,2.22]$.}
\end{figure}

We recall the experimental setup and procedure used to obtain the dispersion relation of in-plane modes by Delory \emph{et al.}~\cite{delory2023}.
A square plate made of a soft elastomer, Ecoflex OO-30, with size $L_0 = \qty{60}{\centi\meter}$ and thickness $h_0 = \qty{3}{\milli\meter}$ is held vertically using clamps at its top and bottom edges (figure~\ref{fig:1}a).
By setting the distance between the clamps $L=\lambda L_0$, we impose a nearly uniaxial deformation described by the principal stretches $(\lambda_1 = \lambda,\lambda_2,\lambda_3)$.
Tracking the position of four markers placed at the center of the plate allow to measure the in-plane stretch ratios.
We find that the imposed deformation is not exactly uniaxial, $\lambda_3 = \lambda^{-0.41}$, as the clamps prevent stretching in the $\boldsymbol{e}_3$ direction.
We also emphasize that the vertical plate is not in its reference configuration in the absence of an imposed elongation. 
Its own weight generates a small non-homogeneous deformation, corresponding to $\lambda = 1.03$ at the center of the plate, that we take into account in this contribution.
A line source driven by a shaker excites monochromatic elastic waves with frequencies $f$ ranging from $\qtyrange{50}{300}{\hertz}$ in the directions $\boldsymbol{e}_1$, parallel to the direction of elongation, or $\boldsymbol{e}_3$, perpendicular to the direction of elongation for values of $\lambda$ spanning the range from 1.03 to 2.27.
The in-plane displacement $(u_1, u_3)$ associated to the propagation of the small-amplitude waves is recorded using stroboscopic imaging in combination with a Digital Image Correlation algorithm~\cite{delory2022}.
For additional details on the experimental procedure to measure the in-plane guided modes, we refer the reader to~\cite{delory2023,delory2022}.

Performing spatial Fourier transforms on each measured monochromatic displacement field enables to obtain the dispersion relations of waves propagating in the parallel and perpendicular directions for $\lambda \in [1.03,2.27]$ (figure \ref{fig:1}b-c).
We indeed observe that two modes propagate, and identify the $S\!H_0$ and $S_0$ modes from their expected velocity in the long-wavelength limit in the absence of pre-stress: $V_T$ and $V_P = 2V_T$, respectively~\cite{royer1999,laurent2020plane}.
As $\lambda$ increases, wave propagation becomes anisotropic: 
parallel to the imposed stretch, both phase velocities increase with $\lambda$, similarly to a string in the tension regime, while in the perpendicular direction, the velocities of the $S\!H_0$ and $S_0$ modes are barely affected.

\subsection{Out-of-plane mode}
We use a similar setup to measure the dispersion curves of the flexural mode (figure \ref{fig:1}d).
We prepare an Ecoflex OO-30 plate with dimensions $L_0 = \qty{60}{\centi\meter}$ and $h_0 = \qty{2.5}{\milli\meter}$, and follow the same procedure to impose an almost uniaxial homogeneous deformation. 
Similarly, we track markers to measure the in-plane principal stretches and obtain $\lambda_3 = \lambda^{-0.39}$.
After the static uniaxial deformation is imposed, a shaker drives a line source in the $\boldsymbol{e}_2$ direction, creating small amplitude out-of-plane waves. The source generates a quadratic chirp signal with instantaneous frequency varying from $\qtyrange{200}{1}{\hertz}$ in $0.7$ seconds.
We record the deflection of a laser sheet projected in oblique incidence and passing through the plate's center and oriented along the $\boldsymbol{e}_1$ or $\boldsymbol{e}_3$ direction with a CCD camera working at an acquisition frequency of $\qty{1}{\kilo\hertz}$.
The out-of-plane displacement $u_2(x_i,t)$, with $i=1,3$, is directly proportional to the in-plane deflection of the laser line, which we extract from the images.
We restrict the measurement region of interest (ROI) to a $\qty{23.5}{\centi\meter} \times \qty{23.5}{\centi\meter}$ square at the center of the plate where the influence of the clamped boundary conditions at the top and bottoms edges of the plate vanishes and the deformation is homogeneous~\cite{delory2023}.

We then obtain the dispersion curves of out-of-plane waves propagating in directions 1 and 3 by performing the two-dimensional Fourier transform of $u_2(x_i,t)$ and by extracting maxima in the $(k_{x_i}, f)$ space.
In figure~\ref{fig:1}(e-f), we represent the dispersion relations of the $A_0$ mode parallel and perpendicular to the direction of elongation for values of the static stretch $\lambda$ ranging from 1.11 to 2.22.
Even for the lowest imposed stretch, we observe that wave propagation is anisotropic. 
Indeed, the $A_0$ mode is very sensitive to the imposed elongation.
It quickly loses its expected parabolic shape in the long-wavelength regime to become non-dispersive in the direction of elongation for the largest imposed values of $\lambda$.
Perpendicular to the direction of propagation, the dispersion curves are less dramatically affected by the imposed stretch, yet they are more markedly influenced than for the two in-plane modes. 

Having evidenced how a large uniaxial deformation affects the propagation of the three zero order modes propagating in a plate, both parallel and perpendicular to the direction of imposed stretch, we turn to the modelling of nearly incompressible hyperelastic materials.

\section{Material modelling}
\label{sec:3}
\subsection{Non-linear elasticity for hyperelastic materials}
We are interested in  nearly incompressible isotropic hyperelastic materials whose mechanical behavior is described by a strain-energy density $W$. 
The constitutive law is a function of the first three principal invariants of the left Cauchy-Green tensor $\boldsymbol{B} = \boldsymbol{F}\cdot \boldsymbol{F}^T$
\begin{align*}
    &I_1 = \mathrm{tr}(\boldsymbol{B}) = \lambda_1^2 + \lambda_2^2 +\lambda_3^2, \\
    &I_2 = \frac{1}{2}\left(\mathrm{tr}(\boldsymbol{B})^2 - \mathrm{tr}(\boldsymbol{B}^2)\right) = \lambda_2^2\lambda_3^2 + \lambda_1^2\lambda_3^2 + \lambda_1^2\lambda_2^2, \\
    &I_3 = \mathrm{det}(\boldsymbol{B}) = \lambda_1^2\lambda_2^2\lambda_3^2 = J^2,
\end{align*}
where $\boldsymbol{F}$ is the deformation gradient and $\lambda_i$ are the principal stretches of the deformation. 

The strain-energy density must also satisfy the following additional constraints~\cite{horgan2004}.
(i) For normalization, we impose that $W$ vanishes in the reference configuration.
(ii) To reflect that a sample cannot be indefinitely expanded or compressed, $W$ should become infinite as $J$ tends to $+\infty$ and $0^+$, respectively.
(iii) We assume that the Cauchy stress vanishes in the reference configuration.
(iv) $W$ must be compatible with the linear elasticity theory so that 
\begin{align}
    \mu_0 &= 2 \left[\frac{\partial W}{\partial I_1} + \frac{\partial W}{\partial I_2} \right]_{(I_1=3,I_2=3,I_3=1)}, \\
    \frac{\kappa_0}{4} + \frac{\mu_0}{3} &= \left[\frac{\partial^2 W}{\partial I_1^2} + 4 \frac{\partial^2 W}{\partial I_1 \partial I_2} + 4 \frac{\partial^2 W}{\partial I_2^2}  + 2 \frac{\partial^2 W}{\partial I_1 \partial I_3} + 4 \frac{\partial^2 W}{\partial I_2 \partial I_3} + \frac{\partial^2 W}{\partial I_3^2}\right]_{(3,3,1)},
\end{align}
where $\mu_0$ and $\kappa_0$ are the infinitesimal shear and bulk modulus, respectively.
More generally, we will use strain energy densities compatible with fourth order weakly non-linear elasticity which has been shown to be the minimal model to capture fully non-linear effects~\cite{destrade2017,norris2024}.

We build compressible strain energy density functions by adding a volumetric term to the classical incompressible strain energy density $W_I$. 
We decompose the deformation into an isochoric and a volumetric part, so that $W$ writes
\begin{equation}
    W\left(\bar{I}_1, \bar{I}_2,J\right) = W_{dev}\left(\bar{I}_1, \bar{I}_2\right) + W_{vol}\left(J\right),
\end{equation}
where $\bar{I}_1 = I_1 / J^{2/3}$, $\bar{I}_2 = I_2 / J^{4/3}$, and $W_{vol}$ is the volumetric contribution. Indeed, this expression must satisfy the fact that $W_{dev}=W_{I}$ in the isochoric case when $J^2 = 1$~\cite{horgan2004}. 
In the following, we fix the form of the volumetric contribution
\begin{equation}
    W_{vol} = \frac{\kappa_0}{2}\left(J-1\right)^2.
\end{equation}
The nearly incompressible nature of the materials under consideration is encoded in the ratio $\mu_0/\kappa_0$ which is a small, but non-zero parameter.

\subsection{Fitting the strain energy density to uniaxial data}
\begin{figure}
    \centering
    \includegraphics[width = 12cm]{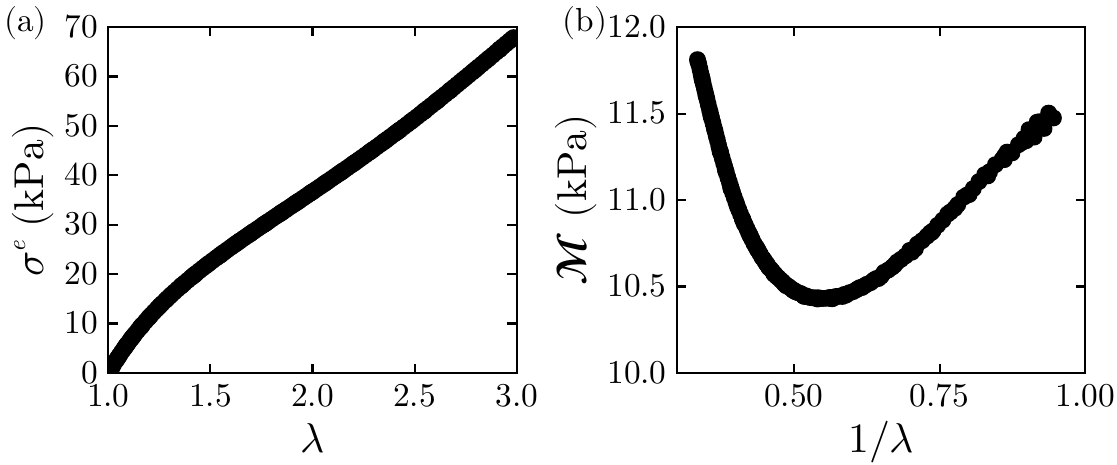}
    \caption{\label{fig:2} Uniaxial extension test of an Ecoflex OO-30 sample. (a) Engineering stress $\sigma^e$ as a function of the stretch ratio in the direction of elongation, $\lambda$. 
    (b) Same data as in (a) represented in the Mooney-space (equation \ref{eq:mooneyspace}).}
\end{figure}

We recall the interpretation of the data gathered during a traction test and the fitting procedure detailed in Destrade \emph{et al.}~\cite{destrade2017}.
In such a test, a uniaxial elongation, corresponding to $\lambda_1=\lambda$, is imposed and the force exerted on the sample in the direction of elongation is recorded. 
By assuming that the deformation is homogeneous and that the material is incompressible, we obtain the principal stretches as $(\lambda, 1 / \sqrt{\lambda}, 1 / \sqrt{\lambda})$, and use the incompressible form of $W$.
In figure~\ref{fig:2}(a), we display the data gathered while performing such a test on an Ecoflex OO-30 sample.
We represent the engineering stress in the direction of elongation $\sigma^e$, that is the force divided by the cross-sectional area of the sample in the reference configuration, as a function of $\lambda$.
In this idealized testing configuration, it is straightforward to compute $\sigma^e$ from the strain energy density
\begin{equation}
    \sigma^e = \frac{\partial W_I}{\partial \lambda} = 2 \left(\lambda - \frac{1}{\lambda^2}\right)\left(\frac{\partial W_I}{\partial I_1} + \frac{1}{\lambda}\frac{\partial W_I}{\partial I_2}\right).
\end{equation}
At this point, it is convenient to introduce the Mooney transform 
\begin{equation}
    \mathcal{M}\left(\xi\right) = \frac{\partial W_I}{\partial I_1} + \xi \frac{\partial W_I}{\partial I_2}, \; \quad \text{where} \;\;\; \mathcal{M} = \frac{\sigma^e}{2(\lambda - 1 / \lambda^2)} \;\;\; \text{and} \;\;\;\xi = \frac{1}{\lambda}.
    \label{eq:mooneyspace}
\end{equation}
We represent $\mathcal{M}$ as a function of $1/\lambda$, the so-called Mooney plot (figure \ref{fig:2}b), and observe that the Mooney transform emphasizes the range of stretch $\lambda \in [1, 2]$.
The Mooney plot crucially allows to distinguish three successive stretch regimes: the small to moderate strain regime, where $\mathcal{M}$ decreases as $\lambda$ increases (here typically for $\xi$ from 1 down to 0.6), the strain-hardening regime corresponding to the upturn in the Mooney plot (for $\xi$ in between 0.6 and 0.4), and the large deformation regime in which the stress rises dramatically at low values of $\frac{1}{\lambda}$~\cite{destrade2017}.
We also emphasize that the initial decrease in the Mooney plot already evidences the limitations of the neo-Hookean model,
\begin{equation}
    W_I\left(I_1\right) = \frac{\mu_0}{2}\left(I_1 - 3\right),
\end{equation}
which predicts a constant value $\mathcal{M} = \mu_0/2$ (equation \eqref{eq:mooneyspace}) and thus must be complemented by an additional $C_2$ term, dependent on $I_2$, even at small elongations~\cite{destrade2017}.

We now focus on matching the traction test data to mathematical models of $W_I$.
We look for the set of parameters $\boldsymbol{p}$, involved in the definition of $W_I$, that best fits the $m$ experimental data points obtained during the traction test.
To find this set of parameters, we follow Destrade \emph{et al.}~\cite{destrade2017} and choose to minimize the 2-norm of the relative residuals $r_i(\boldsymbol{p})$ in the Mooney-space
\begin{equation}
    {\lVert\boldsymbol{r}(\boldsymbol{p})\rVert_2}^2 =   \sum_{i=1}^m r_i(\boldsymbol{p})^2= \sum_{i=1}^m \left(\frac{\mathcal{M}(\lambda_i, \boldsymbol{p})}{\mathcal{M}_i} - 1\right)^2,
    \label{eq:goodnessfit}
\end{equation}
either using a least-square approach, or, when the problem is non-linear with respect to $\boldsymbol{p}$, using a genetic algorithm to ensure we obtain a global minimum~\cite{hansen2016}.
We note that this approach has the advantage to ensure that the fit parameters are the same regardless of the stress, that is we could have equivalently minimized the norms of the relative residuals in the engineering- or Cauchy-space.

\subsection{Guided waves dispersion relations}
We describe how to compute the dispersion relation of small amplitude guided waves propagating in a plate subjected to a large homogeneous deformation from the fitted form of the strain energy density.
These predictions are obtained from the acoustoelastic theory, which describes incremental motions superimposed on the large static deformation~\cite{ogden1997,destrade2007}.
Within this framework, the incremental displacement $\boldsymbol{u}$ obeys a wave-like equation
\begin{equation}
    C_{jikl}^0\left(\lambda_1, \lambda_2, \lambda_3, \boldsymbol{p}\right) \frac{\partial^2 u_k}{\partial x_j \partial x_l} = \rho \frac{\partial^2 u_i}{\partial t^2},
    \label{eq:waveeq}
\end{equation}
where the fourth order elasticity tensor has been replaced by its push-forward version in the deformed configuration $\boldsymbol{C}^0\left(\lambda_1, \lambda_2, \lambda_3, \boldsymbol{p}\right)$~\cite{ogden1997,destrade2007}.  

\begin{figure}
    \centering
    \includegraphics[width = \textwidth]{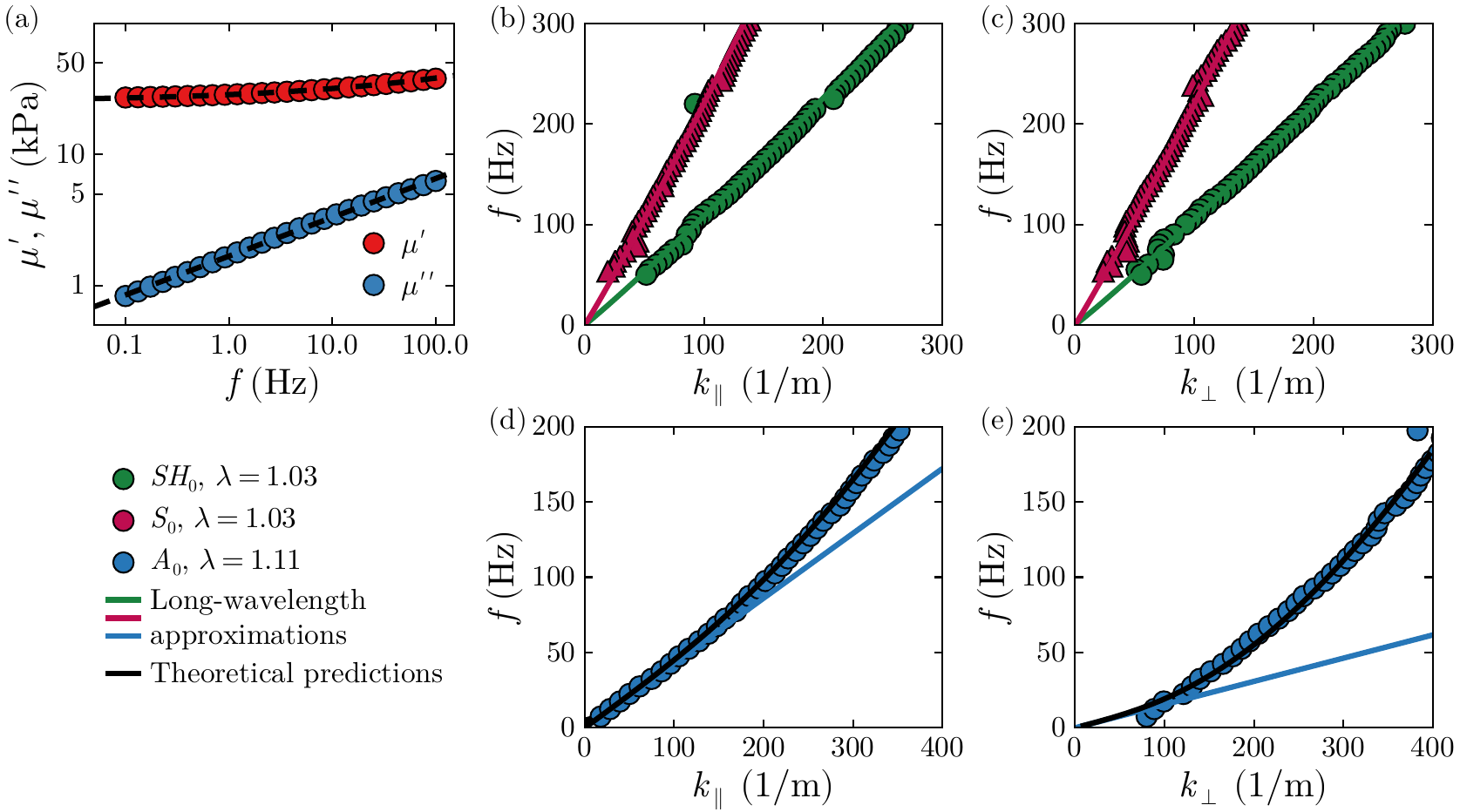}
    \caption{\label{fig:3} (a) Rheological measurement of the complex shear modulus of Ecoflex OO-30. We plot the real and imaginary parts of the shear modulus $\mu'$ and $\mu''$, respectively. The dashed lines show the fit of the fractional Kelvin-Voigt model to the data.
    (b-c) Dispersion relations of the in-plane $S\!H_0$ and $S_0$ modes propagating parallel or perpendicular to the direction of elongation for $\lambda = 1.03$. The solid lines represent the long-wavelength approximations defined in equations \eqref{eq:SH0} and \eqref{eq:S0}.
    (d-e) Dispersion relations of the out-of-plane $A_0$ mode propagating parallel or perpendicular to the direction of elongation for $\lambda = 1.11$. The blue lines stand for the long-wavelength predictions of equation \eqref{eq:A0} and the black lines show the full semi-analytical predictions.}
\end{figure}

An additional difficulty arises for soft elastomers such as those used in our experiments: describing wave propagation requires to take into account their viscoelastic behavior~\cite{yasar2013,liu2014,delory2023}.
According to independent rheological measurements~\cite{lanoy2020}, we model the frequency dependent properties of Ecoflex OO-30 using the fractional derivative Kelvin-Voigt model 
\begin{equation}
    \mu = \mu_0 \left(1 + (\mathrm{i}\omega \tau)^n\right),
    \label{eq:kelvinvoigt}
\end{equation}
where $\mu_0$ is the infinitesimal shear modulus, and $\tau$, $n$ are material parameters~\cite{winter1986,long1996}.
The viscoelastic parameters are obtained by minimizing the relative error between the fractional Kelvin-Voigt model and the real ($\mu'$) and imaginary ($\mu''$) part of the shear modulus measured using a rheometer in plate-plate configuration (figure \ref{fig:3}a).
We find $\mu_0 = \qty{25.3}{\kilo\pascal}$, $\tau = 249 \, \mathrm{\mu s}$ and $n = 0.3$.
We keep the values of $\tau$ and $n$ constant for the remainder of the paper, but we choose to determine $\mu_0$ from the uniaxial extension test which has the advantage to be conducted on a sample prepared in the same conditions as that used in the experiments.
Viscoelasticity is incorporated into the modified elasticity tensor $\boldsymbol{C}^0$ which becomes frequency dependent and now reads
\begin{equation}
    C^0_{jikl}(\lambda_1,\lambda_2,\lambda_3,\boldsymbol{p}, \omega, \beta') = C^0_{jikl}(\lambda_1,\lambda_2,\lambda_3,\boldsymbol{p}) 
    +\mu_0I_{jikl}\left(1 + \beta'\frac{\lambda_i^2+\lambda_j^2-2}{2}\right)(\mathrm{i}\omega\tau)^n,
    \label{eq:elasticitytensor}
\end{equation}
with $I_{jikl} = (\delta_{jk}\delta_{il}+\delta_{jl}\delta_{ik})$, and $\beta'$ is an adjustable parameter~\cite{delory2023, berjamin2025}.
We compute the push-forward of the elasticity tensor for a given choice of mathematical model for the strain energy $W$ using the definitions of its coefficients given in~\cite{delory2024} which are identical to that given in~\cite{ogden1997,destrade2007} with a permutation of the last two indices.
We stress that we must use the compressible form of $W$ to compute this tensor as guided elastic waves involve coupling between longitudinal and shear vertical waves.
We note that taking into account the coupling between viscoelasticity and hyperelasticity results in the introduction of an extra adjustable parameter $\beta'$, which value must be determined from dynamic measurements.

We then proceed to determine the dispersion relation of guided elastic waves propagating in the plate parallel and perpendicular to the direction of imposed elongation $\lambda_1 = \lambda$, with a focus on the shear horizontal, pseudo-longitudinal, and flexural fundamental modes probed in our experiments.
In this situation, we experimentally access only the in-plane principal stretches $\lambda$ and $\lambda_3$.
We thus further take the nearly incompressible limit, $J \to 1$, which in combination with the boundary condition $\sigma_2 = 0$ allows to eliminate the $\lambda_2$ dependence and to get expressions for the coefficients of the tensor $\boldsymbol{C}^0\left(\lambda,\lambda_3,\boldsymbol{p},\omega,\beta'\right)$~\cite{delory2024,kiefer2024_2}.
Guided waves are propagative solutions of equation \eqref{eq:waveeq} that satisfy the boundary conditions
\begin{equation}
    \left.\boldsymbol{e}_2 \cdot \boldsymbol{\sigma}\right\vert_{\pm h/2} = 0,
    \label{eq:waveeqbc}
\end{equation} 
where $h = h_0 \lambda_2 = h_0 \lambda^{-1}\lambda_3^{-1}$. 
For shear horizontal modes, reflections at the boundaries do not involve coupling between different polarisations and the phase velocity of the $S\!H_0$ mode is directly given by the bulk shear wave velocity.
It reads in the directions parallel and perpendicular to the elongation
\begin{subequations}
    \label{eq:SH0}
    \begin{align}
        \begin{minipage}{0.5\linewidth}
            \begin{equation}
                \rho V_{S\!H_0,\parallel}^2 = C^0_{1331}, 
                \label{eq:SH011}
            \end{equation}
        \end{minipage}
        \begin{minipage}{0.5\linewidth}
            \begin{equation}
                \rho V_{S\!H_0,\perp}^2 = C^0_{3113}.
                \label{eq:SH012}
            \end{equation}
        \end{minipage} 
    \end{align}
\end{subequations}
Note that in the absence of prestress these two quantities are equal owing to the symmetries of the elasticity tensor, a fact that is implicitly assumed when adopting the Voigt notation. 

For the pseudo-longitudinal $S_0$ and the flexural $A_0$ Lamb modes, the coupling between shear vertical and longitudinal polarisations at the boundaries complexifies the obtention of the spectrum.
Nevertheless, Rogerson \emph{et al.}~\cite{rogerson2009} derived long-wavelength approximations for the phase velocities of the $S_0$ modes 
\begin{subequations}
    \label{eq:S0}
    \begin{align}
        \begin{minipage}{0.5\linewidth}
            \begin{equation}
                \rho V_{S_0,\parallel}^2 = C^0_{1111} - \frac{\left.C^0_{1122}\right.^2}{C^0_{2222}},
                \label{eq:S011}
            \end{equation}
        \end{minipage}
        \begin{minipage}{0.5\linewidth}
            \begin{equation}
                \rho V_{S_0,\perp}^2 = C^0_{3333} - \frac{\left.C^0_{3322}\right.^2}{C^0_{2222}},
                \label{eq:S012}
            \end{equation}
        \end{minipage}
    \end{align}
\end{subequations}
and the $A_0$ modes 
\begin{subequations}\label{eq:A0}
    \begin{align}
    \begin{minipage}{0.5\linewidth}
        \begin{equation}
            \rho V_{A_0,\parallel}^2 = C^0_{1221} - C^0_{2112},
            \label{eq:A011}
        \end{equation}
    \end{minipage}
    \begin{minipage}{0.5\linewidth}
        \begin{equation}
            \rho V_{A_0,\perp}^2 = C^0_{3223} - C^0_{2332}.
            \label{eq:A012}
        \end{equation}
    \end{minipage}
    \end{align}
\end{subequations}
Again, in the absence of initial deformation, the symmetries of the elasticity tensor yield a null velocity for the $A_0$ mode, which corresponds to the well known parabolic nature of its dispersion relation.

Anticipating later discussions, we evidence the relationship between the phase velocities in the long-wavelength limit and the strain energy density using the formulas given in~\cite{delory2024}.
For the $S\!H_0$ mode, we find
\begin{subequations}
    \label{eq:SH02} 
    \begin{align}
        \rho V_{S\!H_0,\parallel}^2 = \frac{\lambda^2}{\lambda^2 - \lambda_3^2}\frac{1}{J}\left(\lambda \frac{\partial W}{\partial \lambda} - \lambda_3 \frac{\partial W}{\partial \lambda_3}\right) + \mu_0 \left(1 + \beta' \frac{\lambda^2+\lambda_3^2 - 2}{2}\right)(\mathrm{i}\omega\tau)^n, \\
        \rho V_{S\!H_0,\perp}^2 = \frac{\lambda_3^2}{\lambda^2 - \lambda_3^2}\frac{1}{J}\left(\lambda \frac{\partial W}{\partial \lambda} - \lambda_3 \frac{\partial W}{\partial \lambda_3}\right) + \mu_0 \left(1 + \beta' \frac{\lambda^2+\lambda_3^2 - 2}{2}\right)(\mathrm{i}\omega\tau)^n.
    \end{align}
\end{subequations}
We highlight that subtracting these two velocities, we can recover the state of stress in the plate $\sigma_1 - \sigma_3$ using $\sigma_1 = J^{-1}\lambda \partial W / \partial \lambda$ and $\sigma_3 = J^{-1}\lambda_3 \partial W / \partial \lambda_3$, independent of the material rheology, as noted in~\cite{li2022,zhang2023}.
Similarly, we obtain the velocity of the $A_0$ mode parallel and perpendicular to the direction of propagation. Both phase speeds are independent of the material rheology and thus of $\beta'$, and are related to the stress in the plate by 
\begin{subequations}
    \label{eq:A02}
    \begin{align}
        \begin{minipage}{0.5\linewidth}
            \begin{equation}
                \rho V_{A_0,\parallel}^2 = \sigma_1,
                \label{eq:A021}
            \end{equation}
        \end{minipage}
        \begin{minipage}{0.5\linewidth}
            \begin{equation}
            \rho V_{A_0,\perp}^2 = \sigma_3,
                \label{eq:A022}
            \end{equation}
        \end{minipage}
    \end{align}
\end{subequations}
where we took into account the boundary condition $\sigma_2 = 0$, and considered that the deformation is not exactly uniaxial, \emph{i.e.} $\lambda_3 \neq \lambda^{-1/2}$, as observed in our experiments. 
Linking the phase velocity of the $S_0$ mode to $W$ does not yield an expression that can be easily interpreted. 
However, we emphasize that the velocity depends on second order derivatives of $W$ with respect to $\lambda_i$, that is to the derivative of stress with stretch, unlike for the $S\!H_0$ and $A_0$ modes where the velocities are only related to the stress in the plate.

We discuss the relevance of the above approximations by comparing their predictions with the dispersion relations of the three fundamental modes in the parallel and perpendicular directions for $\lambda = 1.03$ for the in-plane modes, and $\lambda = 1.11$ for the out-of-plane mode (figure \ref{fig:3}b-e).
Here, our goal is not to discriminate different mathematical models for $W$ and we compute the modified elasticity tensor for the Mooney-Rivlin strain energy density~\cite{mooney1940,rivlin1948} which was found to accurately describe their data by  Delory \emph{et al.}~\cite{delory2023}.
As the measurements of the in-plane and out-of-plane modes were conducted on distinct Ecoflex OO-30 sample, we use two sets of parameters $\boldsymbol{p}$ which we extract by fitting the Mooney-Rivlin strain energy density to the uniaxial extension data presented in figure \ref{fig:2} for the out-of-plane case, and in figure~6 of reference~\cite{delory2023} for the in-plane case.
We use the procedure described in section \ref{sec:3} that we will systematically detail in section \ref{sec:4}.
We find that the two sets of parameters are equivalent once normalized by the value of the infinitesimal shear modulus $\mu_0 = \qty{31.1}{\kilo\pascal}$, and $\mu_0 = \qty{23.4}{\kilo\pascal}$ for the in-plane and out-of-plane case, respectively.
The principal stretches $\lambda_1 = \lambda$ and $\lambda_3$  are measured, and we, for now, take the value of the remaining parameter, $\beta' = 0.29$, from the work of Delory \emph{et al.}~\cite{delory2023}.
For both in-plane modes, the long-wavelength approximation (solid lines in figure \ref{fig:3}b-c) accurately captures the dispersion relation in the whole experimental range.
In contrast, for the flexural mode (figure \ref{fig:3}d-e), the approximations are satisfying only for a limited wavenumber range, especially perpendicular to the direction of elongation, as expected for this bending mode which displays a parabolic dispersion relation in the reference configuration.

The inability of equations \eqref{eq:A0} to fully capture the dispersion relation of the flexural mode highlights the need to go beyond long-wavelength predictions.
We do so by searching for non-trivial solutions to the system of equations \eqref{eq:waveeq} and \eqref{eq:waveeqbc} using a semi-analytical method.
After inserting the displacement ansatz $\boldsymbol{u} = \boldsymbol{u}\left(k_{i},x_2,\omega\right)e^{\mathrm{i}(k_{i}x_i-\omega t)}$ with $i = 1,3$, we discretize the problem in the $x_2$ direction using the spectral collocation method~\cite{adamou2004,shen2011}.
We obtain the dispersion relation as pairs $(k_{i}, \omega)$ solution to an algebraic eigenvalue problem where the wavenumber is complex valued, a key to successfully model viscoelastic media~\cite{kiefer2022}.
Our implementation, adapted from~\cite{kiefer2024_2,kiefer2024}, is detailed and available in~\cite{chantelot2026}.
The semi-analytical predictions for the dispersion relation of the $A_0$ mode are shown with a black line on figure \ref{fig:3}(d-e) and exhibit excellent agreement with the experimental measurements.

We now possess all the necessary tools to fit the strain-energy density to uniaxial extension data and to predict the dispersion relations of guided modes propagating in an homogeneously deformed plate characterized by the strain energy $W$, enabling us to perform a comparison with our dynamic measurements.
We highlight that our experiments have the potential to provide new pieces of information over a static test: the pseudo-longitudinal mode is sensitive to the derivative of stress with strain, and the flexural mode gives insight on the stress perpendicular to the direction of elongation.
However, this additional data comes at the expense of involving an extra fit parameter $\beta'$ that accounts for the intrinsic rheology of the material.

\section{Results}
\label{sec:4}
We now compare our experimental measurements to predictions obtained for different constitutive laws fitted on the uniaxial extension data of figure \ref{fig:2}.
We successively address the three stretch regimes evidenced in the Mooney plot to eventually describe the material behavior over the whole range of imposed elongation.

\subsection{Small to moderate strain regime}
\begin{figure}
    \centering
    \includegraphics[width = \textwidth]{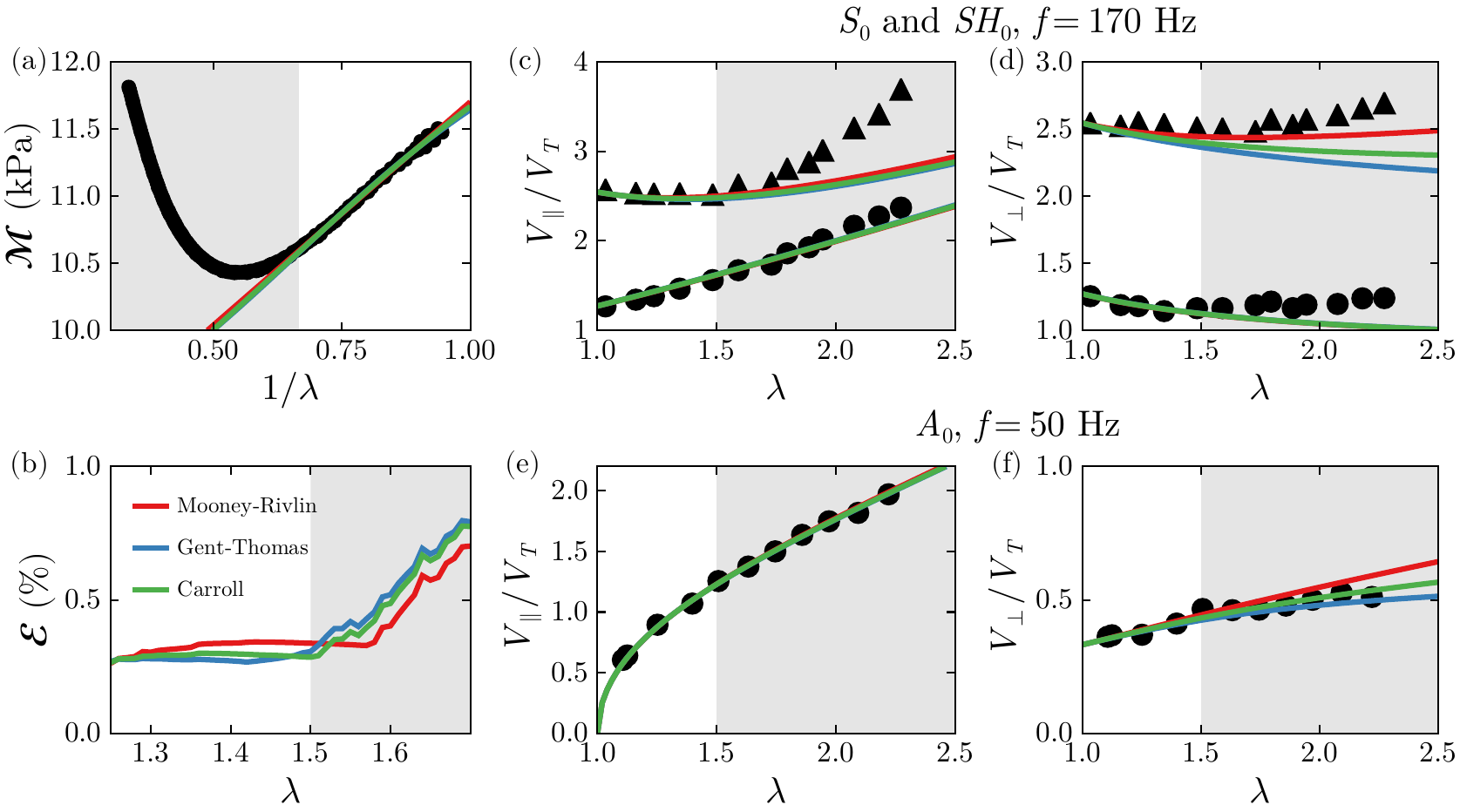}
    \caption{\label{fig:4} (a) Mooney plot for Ecoflex OO-30 showing the fits of the Mooney-Rivlin, Gent-Thomas and Carroll models in the small to moderate stretch regime (solid lines). The greyed-out region evidences the points for which $\lambda > 1.5$ that are excluded from the fitting procedure.
    (b) Maximum relative error $\mathcal{E}$ as a function of the fit endpoint for the three models considered in (a).
    (c-d) Normalized wave velocities for the $S\!H_0$ (dots) and $S_0$ (triangles) modes propagating parallel and perpendicular to the direction of elongation as a function of $\lambda$ for $f = \qty{170}{\hertz}$. The solid lines are long-wavelength predictions computed from equations \eqref{eq:SH0} and \eqref{eq:S0}.
(e-f) Normalized wave velocities in the directions $\boldsymbol{e}_1$ and $\boldsymbol{e}_3$ for the $A_0$ mode for $f = \qty{50}{\hertz}$. Parallel to the direction of propagation, the predictions come from the approximation \eqref{eq:A011} while perpendicular to the direction of propagation we represent the results of the semi-analytical theory.}
\end{figure}
We first focus on the small to moderate strain regime, which corresponds to the linear decrease in the Mooney plot (figure \ref{fig:4}a), where the neo-Hookean model must be augmented by the addition of $C_2$ term dependent on the second invariant $I_2$.
We consider three models that can in principle capture the decrease in the Mooney plot and differ only by the functional form of the $I_2$ term:
\begin{itemize}
\item the Mooney-Rivlin model~\cite{mooney1940}
\begin{equation}
    W_I\left(I_1, I_2\right) = \frac{C_1}{2}\left(I_1 - 3\right) + \frac{C_2}{2}\left(I_2 - 3\right),
\end{equation}
\item the Gent-Thomas model~\cite{gent1958}
\begin{equation}
    W_I\left(I_1, I_2\right) = \frac{C_1}{2}\left(I_1 - 3\right) + \frac{3C_2}{2}\ln \left(\frac{I_2}{3}\right),
\end{equation}
\item the Carroll model~\cite{carroll2011}
\begin{equation}
    W_I\left(I_1, I_2\right) = \frac{C_1}{2}\left(I_1 - 3\right) + \sqrt{3}C_2\left(\sqrt{I_2} - \sqrt{3}\right).
\end{equation}
\end{itemize}
We fit these two-parameter models to the uniaxial data of figure \ref{fig:4}(a) using the goodness of fit metric defined in equation \eqref{eq:goodnessfit} by introducing generalized Mooney transforms~\cite{anssari2022}.
Similarly as equation \eqref{eq:mooneyspace} enables the Mooney-Rivlin model to be represented as a line in the $\left(\mathcal{M}, \xi\right)$ space, these transforms allow to fit the Gent-Thomas and Carroll models using linear regressions.
However, it requires to rigorously determine the extent of the small to moderate stretch regime which we do by recording the maximum of the relative error, 
\begin{equation}
    \mathcal{E} = \max_i \left|r_i\left(\boldsymbol{p^*}\right)\right|,
\end{equation}
while varying the fit endpoint, as proposed by Destrade \emph{et al.}~\cite{destrade2017}.
We display the results in figure~\ref{fig:4}(b).
The maximum relative error rises sharply when the fit endpoint corresponds to a principal elongation $\lambda > 1.5$, which we take as a bound of the small to moderate strain regime.
We then determine the best fit values of the parameters $C_1$ and $C_2$ for each hyperelastic model and report them in table \ref{tb:smalltomoderate}. 
\begin{table}
\centering
\caption{Material parameters $C_1$ and $C_2$ for the Mooney-Rivlin, Gent-Thomas, and Carroll hyperelastic models obtained by fitting the uniaxial extension data for $\lambda < 1.5$. We also report the values of the shear modulus $\mu_0 = C_1 + C_2$, and of the parameter $\beta'$.}
\label{tb:smalltomoderate}
\begin{tabular}[]{l c c c c}
    \hline
    Model & $C_1$ (kPa) & $C_2$ (kPa)& $\mu_0$ (kPa) & $\beta'$ \\
    \hline
    Mooney-Rivlin & $16.7$ & $6.6$ & $23.4$ & $0.0$ \\
    Gent-Thomas & $18.2$ & $5.1$ & 23.3  & $0.0$ \\
    Carroll & $17.6$ & $5.7$ & 23.3 & $0.0$ \\
    \hline
\end{tabular}
\end{table}
The sets of parameters describing our two distinct Ecoflex OO-30 samples being equivalent once normalized by the shear modulus $\mu_0$, we give here and in the following only the parameters obtained by fitting the data of figure \ref{fig:2}.
We also represent the best fit curves in figure \ref{fig:4}(a) and observe that all three models excellently capture the decrease in the Mooney plot.
As already noted~\cite{carroll2011,destrade2017}, the uniaxial extension test does not allow to discriminate the form of the $C_2$ term.

We then compute the modified elasticity tensor $\boldsymbol{C}^0(\lambda,\lambda_3,\boldsymbol{p},\omega,\beta')$, and obtain the dispersion relations of the three fundamental guided modes.
To evidence the influence of imposed stretch, we compare the predicted dispersion relations to the data of figure \ref{fig:1} at a fixed frequency by representing the phase velocity parallel and perpendicular to the direction of imposed stretch as a function of $\lambda$ for each mode (figure \ref{fig:4}c-f).
We choose an intermediate frequency for the in-plane modes, $f = \qty{170}{\hertz}$, where our measurements are the most reliable, and we choose $f = \qty{50}{\hertz}$ for the out-of-plane mode as it is more sensitive to stretch in the long-wavelength regime.
This choice of frequency is arbitrary and we provide an interactive tool to explore the data and predictions at all frequencies~\cite{chantelot2026_2}.
Using this tool, we verify that the velocity of the $A_0$ mode parallel to the direction of elongation, for which most data lies in the long-wavelength limit, is independent of frequency and thus of the material's rheological behavior and of $\beta'$, as predicted by equation~\eqref{eq:A021}.
To perform a quantitative comparison between the measured and predicted values, we determine $\beta'$ for each hyperelastic model by minimizing the relative error between the measured and predicted phase velocities of the $S\!H_0$ and $S_0$ modes.
The values of $\beta'$ are reported in table~\ref{tb:smalltomoderate}.
We find $\beta'=0$ for all three models, in contrast with the non-zero value reported by  Delory \emph{et al.}~\cite{delory2023} who considered larger values of $\lambda$ in the fit procedure.
This suggests that viscous effects are not coupled to prestress in the small to moderate elongation regime (see equation~\eqref{eq:elasticitytensor}).

We represent the predicted phase velocities with solid lines in figure \ref{fig:4}(c-f).
For the in-plane modes, the predictions are computed from the long-wavelength approximations of equations \eqref{eq:SH0} and \eqref{eq:S0}, while for the $A_0$ mode we use the long-wavelength prediction, equation \eqref{eq:A011}, parallel to the direction of elongation and the semi-analytical prediction in the perpendicular direction.
First, the three models are able to qualitatively reproduce the measured phase velocities for all modes and propagation directions in the small to moderate strain regime.
Moreover, the predictions for the $S\!H_0$ and $A_0$ modes in the direction $\boldsymbol{e}_1$, and to a lesser extent for the $A_0$ mode in direction $\boldsymbol{e}_3$, remain satisfying well outside the range of validity of these weakly non-linear models. 
In contrast, we observe marked deviations at large stretch for the $S_0$ mode, and for the $S\!H_0$ mode in the perpendicular direction.
Second, unlike in the Mooney plot where all models fit equally well the data, the predicted phase velocities for the $S_0$ mode propagating perpendicular to the direction of imposed stretch (black triangles, figure \ref{fig:4}d) significantly differ even when $\lambda < 1.5$, with the Mooney-Rivlin model capturing better the evolution of the phase velocity with $\lambda$.
The fact that the predictions differ can be qualitatively understood by considering the dependence of the phase velocities on $W$.
For the shear horizontal and flexural modes, $V$ is directly related to the stress in the sample, that is to derivatives of $W$ with the principal stretches (equations~\eqref{eq:SH02} and~\eqref{eq:A02}), a quantity that all models fit equally well in the Mooney plot.
In contrast, $V$ depends on second order derivatives of the strain energy density with the principal stretches for the pseudo-longitudinal mode, amplifying minute differences between the three models. 

To summarize, in the small to moderate strain regime, all three hyperelastic models provide an excellent fit of the uniaxial extension data, and satisfyingly capture the dispersion of guided waves propagating in the plate sample.
Yet, we observe a small but significant difference between the predictions for pseudo-longitudinal waves propagating perpendicular to the direction of elongation.
This piece of information allows to discriminate the three models, that is to obtain unique insight on the form of the $I_2$ contribution to the strain-energy density.
In the following, we tackle larger imposed elongations with the goal to identify if this trend persists.

\subsection{Strain-hardening regime}
\begin{figure}
    \centering
    \includegraphics[width = \textwidth]{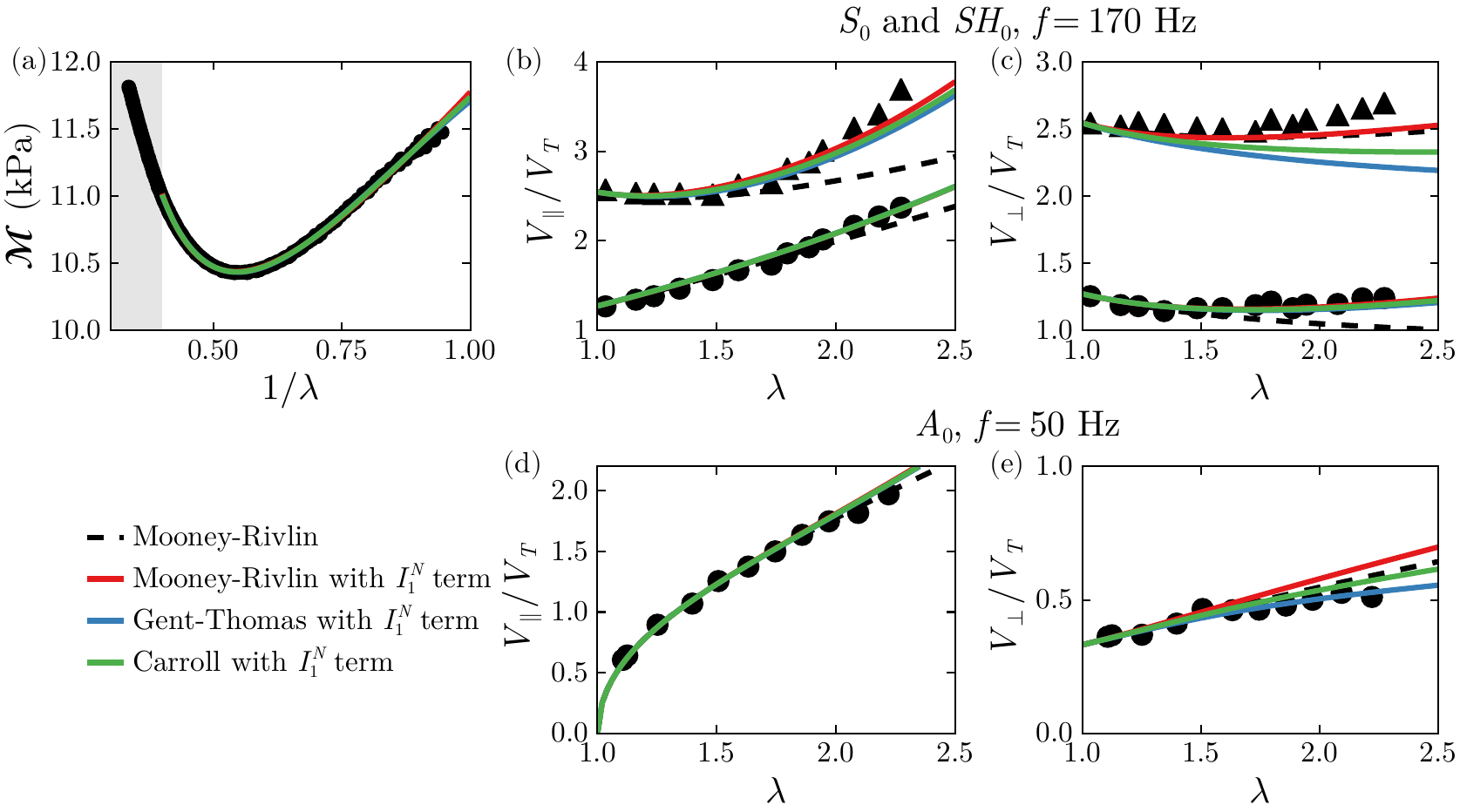}
    \caption{\label{fig:5} (a) Uniaxial elongation test for Ecoflex OO-30 represented in the Mooney space. The solid lines are the fits of equation \eqref{eq:SHI1} for $\lambda < 2.5$ when considering functional forms of the $I_2$ term corresponding to the Mooney-Rivlin, Gent-Thomas, and Carroll models. 
    (b-c) Normalized wave velocities for the $S\!H_0$ (dots) and $S_0$ (triangles) modes propagating parallel and perpendicular to the direction of elongation as a function of $\lambda$ for $f = \qty{170}{\hertz}$. The solid lines are long-wavelength predictions computed from equations \eqref{eq:SH0} and \eqref{eq:S0}.
    (d-e) Normalized wave velocities in the directions $\boldsymbol{e}_1$ and $\boldsymbol{e}_3$ for the $A_0$ mode. Parallel to the direction of propagation, the predictions come from the approximation \eqref{eq:A011}, while perpendicular to the direction of propagation we represent the results of the semi-analytical theory.}
\end{figure}

We move on to the strain-hardening regime which corresponds to the upturn in the Mooney plot~\ref{fig:5}(a). 
Following the methodical approach of Destrade \emph{et al.}~\cite{destrade2017}, we add a power-law term to the previously considered hyperelastic models to account for strain-hardening~\cite{lopez2010}
\begin{equation}
    W_I\left(I_1,I_2\right) = \frac{C_1}{2}\left(I_1 - 3\right) + C_2f(I_2) + \frac{3^{1-N}C_3}{2N}\left(I_1^N - 3^N\right),
    \label{eq:SHI1}
\end{equation}
where $C_3$ and $N$ are constants, and $f\left(I_2\right)$ denotes the functional dependence of $W$ on $I_2$ for the Mooney-Rivlin, Gent-Thomas, and Carroll models.
As $\lambda$ increases, equation \eqref{eq:SHI1} indicates that $W$ is proportional to $\lambda^{2(N-1)}$.
This asymptotic behavior corresponds to $\mathcal{M} \propto \left(1/\lambda\right)^{-2(N-1)}$, so that modelling the upturn in the Mooney plot requires $N>1$.

We choose empirically the extent of the fit, $\lambda \in [1, 2.5]$, making sure that we consider values of $\lambda$ large enough to capture the fact that $\mathcal{M}$ has gone through a minimum.
Finding the set of parameters that best adjusts equation \eqref{eq:SHI1} to the simple elongation test is a non-linear fitting procedure. 
In such a case, the outcome of the fit can be affected by the choice of initial conditions, with multiple local minima or even multiple minima with a similar level of error~\cite{ogden2004}.
In order to remove the dependence on the initial guess, we vary $N$ and perform a linear least-square fit on the parameters $C_1$, $C_2$ and $C_3$ while recording the 2-norm of the relative residuals. The optimal set of parameters $\boldsymbol{p}$ is obtained for the value of $N$ that minimizes this metric~\cite{destrade2017}.
We plot the fits of the three hyperelastic models as solid lines on figure~\ref{fig:5}(a).
All models are in very good agreement with the data and their fitted forms almost superimpose over the whole range of stretches $1 < \lambda < 2.5$, showing the importance of the added term to adjust the upturn of $\mathcal{M}$. Again, the Mooney plot alone is insufficient to discriminate between the models.
We report  in table \ref{tb:strainhardening} the associated set of parameters, and the value of the infinitesimal shear modulus for the three models described by equation \eqref{eq:SHI1}. 
\begin{table}
\centering
\caption{Material parameters $C_1$, $C_2$, $C_3$ and $N$ obtained by fitting the uniaxial extension data to equations $\eqref{eq:SHI1}$ for $\lambda < 2.5$. We also report the values of the infinitesimal shear modulus $\mu_0 = C_1 + C_2 + C_3$, of $\beta'$, and of the maximum of the relative errors between predicted and measured velocities across all modes and propagation directions $\mathcal{E}_V$.}
\label{tb:strainhardening}
\begin{tabular}[]{l c c c c c c c}
    \hline
    Model & $C_1$ (kPa) & $C_2$ (kPa) & $C_3$ (kPa) & N &$\mu_0$ (kPa) & $\beta'$ & $\mathcal{E}_V$ (\unit{\percent})\\
    \hline
    Mooney-Rivlin $+I_1^N$ & $14.3$ & $8.1$ & $1.1$ & $2.6$ & $23.5$ & $0.32$ & $23.7$\\
    Gent-Thomas $+I_1^N$& $15.3$ & $6.3$ & $1.9$ & $2.2$ & $23.4$ & $0.26$ & $16.9$\\
    Carroll $+I_1^N$& $14.7$ & $7.1$ & $1.6$ & $2.3$ & $23.5$ & $0.29$ & $12.7$\\
    \hline
\end{tabular}
\end{table}
The obtained values of $\mu_0$ are in very good agreement with that reported in table~\ref{tb:smalltomoderate} for the small to moderate strain regime. 
We also note that when $f\left(I_2\right)$ is given by the Mooney-Rivlin model, we find a value of $N$ just above the maximal reasonable value of 2.5 empirically introduced by Destrade \emph{et al.}~\cite{destrade2017} and refer the reader to this reference for a discussion on the admissible range of values for $N$. 

Moving to our elastic wave measurements, as in the small to moderate strain regime, we minimize the relative error between the predictions for the phase velocities of the in-plane modes, stemming from the fitted form of the strain-energy densities \eqref{eq:SHI1}, and the experimental measurements at fixed frequency to find the value of $\beta'$ for each model. 
We report the obtained values in table \ref{tb:strainhardening} and remark that they are now comparable to that given by Delory \emph{et al.}~\cite{delory2023}.
When considering elongations beyond the small to moderate strain regime, viscoelasticity and prestress indeed become coupled. 
Importantly, the value of $\beta'$ is  weakly affected by the choice of constitutive law so that our conclusions will not be influenced by the determination of this extra parameter.

We compare the predicted phase velocities (solid lines) to the experimental data in figures~\ref{fig:5}(b\textendash e).
Using the Mooney-Rivlin model predictions as a baseline (black dashed line), we immediately notice that overall the addition of the power-law term significantly improves the ability of the models to describe the experimental data at large $\lambda$, and that the additional strain-hardening term has a similar influence on the behavior of all three strain-energy densities.
Interestingly, unlike in the small to moderate strain regime, the deviations are not limited to the pseudo-longitudinal mode, which probes the derivatives of stress with stretch, but also appear for the flexural mode perpendicular to the direction of elongation.
Qualitatively, the Mooney-Rivlin form of the $C_2$ term provides the best match to the variation of the phase velocity with $\lambda$ for the $S_0$ mode, while the Gent-Thomas form provides the best agreement for the $A_0$ mode.
Quantitatively, we choose to record the maximum of the relative errors between predicted and measured velocities across all modes and propagation directions, that we denote $\mathcal{E}_V$ (table~\ref{tb:strainhardening}).
The Gent-Thomas and Carroll models perform very similarly and exhibit maximal error for the $S_0$ mode perpendicular to the direction of propagation.
In contrast, the largest relative error for the Mooney-Rivlin predictions occurs for the out-of-plane mode in direction $\boldsymbol{e}_3$ with $\mathcal{E}_V = \qty{23.7}{\percent}$.
This behaviour can be understood by recalling that, in the long-wavelength limit, the velocity of the $A_0$ mode in direction $\boldsymbol{e}_3$ is related to the stress $\sigma_3$. 
Measuring the velocity of the flexural mode in the perpendicular direction thus highlights that we do not impose a true uniaxial deformation, \emph{i.e.} $\lambda_3 \neq 1 / \sqrt{\lambda}$.
The stress is biaxial, a deformation class for which the Mooney-Rivlin constitutive law is known to show large error in static tests~\cite{treloar2005}, in agreement with our observations.
Note that if we had imposed a strictly uniaxial deformation, the $A_0$ mode would not be discriminating.
Ignoring this mode in the error calculation, we indeed find that the Mooney-Rivlin form of the $C_2$ term gives the lowest relative error with $\mathcal{E}_V = \qty{7.0}{\percent}$ for the $S_0$ mode perpendicular to the direction of elongation.

The addition of a strain-hardening term to the Mooney-Rivlin, Gent-Thomas and Carroll models extends their ability to predict the dispersion of guided waves to the largest elongations investigated.
As in the previous regime, comparing measured and predicted phase velocities for the three lowest order guided modes allows to distinguish constitutive laws that perform equivalently in the Mooney space.
The data suggest that $I_2$ dependences of the Carroll or Gent-Thomas form perform equally well, while the Mooney-Rivlin form leads to a higher relative error for the out-of-plane mode perpendicular to the direction of elongation.
However, adding a strain-hardening term comes at the expense of introducing two additional fit parameters.
In the following, we explore whether simpler generalized neo-Hookean constitutive laws that allow to describe the full simple tension curve, going beyond the upturn, with only three adjustable constants can also capture our dynamic measurements.

\subsection{Large deformation regime}
We finally turn to hyperelastic models that allow to describe the material behavior over the whole range of imposed stretch.
Modelling the stiffening that occurs in the large deformation regime requires to identify the limiting behavior of polymer chains, which is reflected in the singular behavior of $W$ at large $\lambda$.
We consider two families of phenomenological strain-energy density functions that mimic the large stretch behavior of two different microscopical models.
The Gent model~\cite{gent1996}
\begin{equation}
    W_I(I_1,I_2) = -\frac{C_1C_3}{2}\ln\left(1 - \frac{I_1-3}{C_3}\right) + C_2 f\left(I_2\right),
    \label{eq:Gent}
\end{equation}
which represents the freely jointed chain model~\cite{horgan2002}, and the Dobrynin \& Carrillo model~\cite{dobrynin2011}
\begin{equation}
    W_I(I_1,I_2) = \frac{C_1}{6}\left(I_1-3 + \frac{2C_3}{1 - \left(I_1 -3\right)/C_3} - 2C_3\right) + C_2f\left(I_2\right)
    \label{eq:DC}
\end{equation}
which corresponds to the worm-like chain model~\cite{puglisi2016}.
Note that following the methodical approach of Destrade \emph{et al.}, we complemented each generalized neo-Hookean model with a function of the second invariant that can take the Mooney-Rivlin, Gent-Thomas or Carroll form.

\begin{figure}
    \centering
    \includegraphics[width = \textwidth]{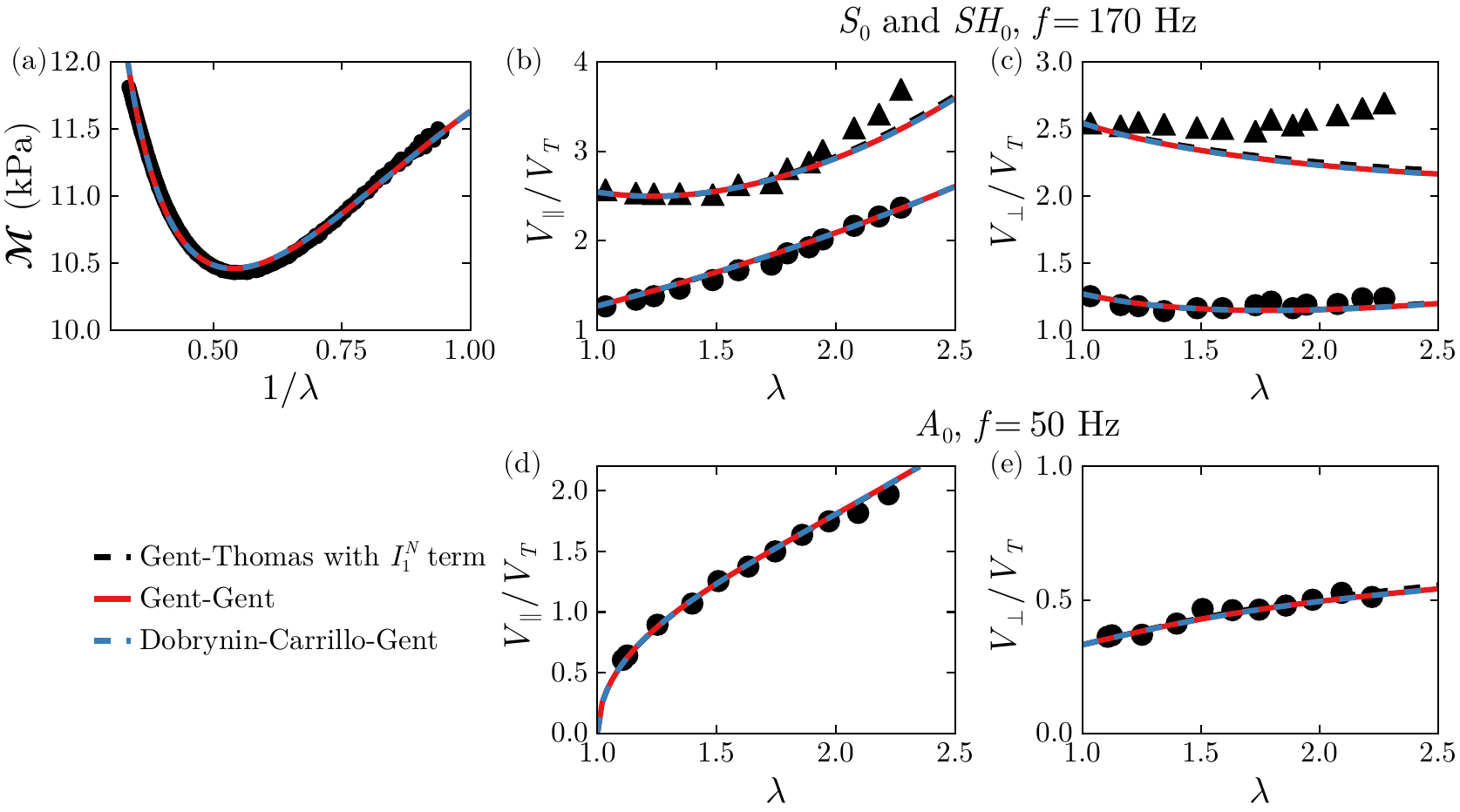}
    \caption{\label{fig:6} (a) Fits of the Gent model (solid line) and of the Dobrynin \& Carrillo model (dashed line) augmented by a $C_2$ term of the Gent-Thomas form to the simple tension test for Ecoflex OO-30.
    (b-c) Normalized wave velocities for the $S\!H_0$ (dots) and $S_0$ (triangles) modes propagating parallel and perpendicular to the direction of elongation as a function of $\lambda$ for $f = \qty{170}{\hertz}$. The lines are long-wavelength predictions computed from equations \eqref{eq:SH0} and \eqref{eq:S0}.
(d-e) Normalized wave velocities in the directions $\boldsymbol{e}_1$ and $\boldsymbol{e}_3$ for the flexural mode at $f = \qty{50}{\hertz}$. Parallel to the direction of propagation, the predictions come from the approximation \eqref{eq:A011}, while perpendicular to the direction of propagation we represent the results of the semi-analytical theory.}
\end{figure}

We focus on the micro-mechanically motivated Gent-Thomas form of the $I_2$ dependent term as we identified that adding a stiffening contribution at large elongation has a similar effect on all models.
As the fit procedure is non-linear, we find the set of parameters $\boldsymbol{p}$ that best describes the uniaxial extension data using a genetic algorithm~\cite{hansen2016}, which we initialise with the values of $C_1$ and $C_2$ reported in table \ref{tb:smalltomoderate} for the Gent-Thomas model, and with $C_3 = 10$. 
The values of the three constants are given in table \ref{tb:largedef}, and we plot the fits of equations \eqref{eq:Gent} and \eqref{eq:DC} to the simple extension test on figure \ref{fig:6}(a).
The fitted forms of $W$ excellently describe the data in the Mooney-space over the whole range of imposed stretch, and we recover a value of the infinitesimal shear modulus in agreement with that previously obtained.

\begin{table}
\centering
\caption{Material parameters $C_1$, $C_2$, and $C_3$ obtained by fitting the uniaxial extension data to equations \eqref{eq:Gent} and \eqref{eq:DC} using the Gent-Thomas form of $f\left(I_2\right)$. We also report the values of the infinitesimal shear modulus $\mu_0 = C_1 + C_2$, and of $\beta'$.}
\label{tb:largedef}
\begin{tabular}[]{l c c c c c}
    \hline
    Model & $C_1$ (kPa) & $C_2$ (kPa) & $C_3$ &$\mu_0$ (kPa) & $\beta'$ \\
    \hline
    Gent-Gent & $17.6$ & $5.6$ & $28.4$ & $23.3$ & $0.25$ \\
    Dobrynin-Carrillo-Gent & $17.7$ & $5.6$ & $38.9$ & $23.3$ & $0.25$ \\
    \hline
\end{tabular}
\end{table}

We compute the modified elasticity tensor $\boldsymbol{C}^0\!\!\left(\lambda, \lambda_3, \boldsymbol{p}, \omega, \beta'\right)$ allowing us to access the dispersion relations of guided waves propagating in the plate.
We give the value of $\beta'$ determined by minimizing the relative error between the measurement and predictions for the phase velocities of the in-plane modes at $f=\qty{170}{\hertz}$ in table~\ref{tb:largedef}, and represent the predictions in figures~\ref{fig:6}(b\textendash e) (solid red line and blue dashed line).
First, the forms of $W$ introduced in equations~\eqref{eq:Gent} and~\eqref{eq:DC} enable to accurately predict all phase velocities for all values of $\lambda$ probed in this study , and perform equivalently as the Gent-Thomas model augmented by an $I_1^N$ term (black dashed line).
Second, the phase velocities deduced from both models are undistinguishable for all three fundamentals modes propagating in directions $\boldsymbol{e}_1$ and $\boldsymbol{e}_3$.
This remains true even in the limiting extension regime, \emph{i.e.} when $\lambda > 2.5$, although we do not access it experimentally.
We stress that, if we wanted to favor the agreement for the $S_0$ in-plane mode, we could have chosen to use the Mooney-Rivlin form of the $C_2$ term to augment both limiting-chain constitutive laws, without affecting the above observations (see the interactive tool \cite{chantelot2026_2}).

The Gent-Gent~\cite{pucci2002} and Dobrynin \& Carrillo~\cite{dobrynin2011} models complemented by a function of the second invariant allow to describe the uniaxial extension test and the propagation of guided waves over the entire range of imposed elongation.
Both models perform equivalently in static and dynamic measurements: we are not able to discriminate the behavior of polymer chains at large extension.
We emphasize that the Gent-Gent and Dobrynin \& Carrillo models capture the material's behavior in all stretch regimes using only three material parameters.
In that light, the two families of strain energy density functions introduced in this subsection can advantageously replace the four-parameter models used to describe strain-hardening in the previous section. 

\section{Conclusion}
\label{sec:6}
In this work, we measure the dispersion relation of the three zero-order guided modes propagating in a plate undergoing uniaxial extension, both parallel and perpendicular to the direction of imposed stretch.
Combining the material's rheological behaviour with the acoustoelastic theory, we show that these dynamic measurements give access to pieces of information that cannot be obtained from the stress-strain curve recorded in corresponding simple tension tests.
Indeed, the pseudo-longitudinal mode is sensitive to second derivatives of the strain-energy density with stretch, and, in the long-wavelength limit, the flexural mode propagating perpendicular to the direction of elongation is sensitive to the stress in the same direction, evidencing deviations from a purely uniaxial stress state. 
This additional insight comes at the expense of introducing an extra fit parameter which couples the viscoelastic and hyperelastic behaviour of our sample.

Using the methodical approach introduced by Destrade \emph{et al.}~\cite{destrade2017}, we demonstrate by analysing data in the small to moderate strain and in the strain-hardening regimes that guided wave measurements enable to discriminate constitutive laws with different functional dependence on $I_2$ that fit equally well the static curve.
We find that the Gent-Thomas and Carroll forms of the $C_2$ term perform comparably well, and overall provide a better description of the whole dataset than the Mooney-Rivlin form.
However, we remark that if we had imposed a purely uniaxial deformation, the Mooney-Rivlin model would provide the best predictions.
In contrast, the dynamic data do not enable to tell apart the predictions from generalized neo-Hookean models that empirically represent distinct singular strain-energy density behaviour in the limiting-chain regime.
We also note that viscous effects and prestress begin to interact only when considering elongations beyond the small to moderate strain regime, and that the value of the coupling parameter is weakly affected by the choice of constitutive law.

These findings echo the insight gained by performing static experiments across multiple deformation classes.
The Mooney-Rivlin form of the $C_2$ term gives the best predictions except when considering the flexural mode propagating perpendicular to the imposed elongation, that is when our experiments are sensitive to the presence of a small biaxial stress.
This observation corroborates that of Treloar~\cite{treloar2005} who evidenced that the Mooney-Rivlin model fits excellently uniaxial data but gives large errors in equibiaxial tests.
For the $A_0$ mode, the Gent-Thomas form of the $C_2$ term provides the best agreement, consistent with the findings of Ogden \emph{et al.}~\cite{ogden2004} who showed that the Gent-Gent model fitted on uniaxial data performs satisfyingly across a variety of deformation classes.

In conclusion, guided-wave measurements in a plate under uniaxial extension reveal constitutive modelling differences that are indistinguishable at the level of the stress–strain curve.
Our results show that a constitutive law combining a generalized neo-Hookean term with an $I_2$-dependent Gent–Thomas or Carroll contribution, such as the Gent–Gent model~\cite{pucci2002}, provides a consistent description of both simple tension tests and incremental wave propagation, with low relative error.

Two natural extensions of this work can be identified. First, exploiting higher-order guided modes would allow the extraction of richer information from dispersion measurements. This could be achieved, for instance, by replacing the plate geometry considered here with a strip geometry, which supports a larger number of guided modes within the same frequency range~\cite{lanoy2020}.
Second, following the approach commonly used in static tests, incremental wave propagation could be investigated under other underlying static deformations, such as biaxial loading.
Beyond dispersion, measuring wave attenuation may provide additional insight into the coupling between hyperelastic and viscoelastic effects. In the present study, only first-order viscous terms are considered, an approximation that may become inaccurate at large elongations, where the constitutive response is strongly nonlinear.
Finally, a wave-based inversion procedure could be developed to obtain the constitutive law's adjustable parameters directly from the dynamic data.
Overall, all these perspectives suggest that incremental wave propagation may become a powerful tool for the mechanical characterization of soft materials. 

\ack{We thank Florent Dorlot and Simon Yves for their help in setting up the experiments, and Alexandre Delory for insightful discussions. 
We acknowledge funding by the Agence Nationale de la Recherche (ANR), grant ANR-24-CE30-3540.}


\vskip2pc

\bibliographystyle{RS}
\bibliography{bibli}

\end{document}